\documentclass[pre,twocolumn,amsmath,amssymb,floats,superscriptaddress,10pt]{revtex4-2}

\usepackage{graphicx}
\usepackage{dcolumn}
\usepackage{bm}
\usepackage{revsymb}
\usepackage[usenames]{color}
\usepackage{subfigure}
\usepackage{color}
\usepackage{physics}
\usepackage{amsmath}
\usepackage{hyperref}

\usepackage{xcolor}

\begin{document}

\title{Odd Cosserat elasticity in active materials}

\author{Piotr Sur\'owka}
\affiliation{Department of Theoretical Physics, Wroc\l{}aw University of Science and Technology, 50-370 Wroc\l{}aw, Poland}
\affiliation{Max Planck Institute for the Physics of Complex Systems, N\"othnitzer Stra\ss e 38, 01187, Dresden, Germany}
\affiliation{W\"urzburg-Dresden Cluster of Excellence ct.qmat, Germany}

\author{Anton Souslov}
\affiliation{Department  of  Physics,  University  of  Bath,  Claverton  Down,  Bath,  BA2  7AY,  UK}

\author{Frank J\"{u}licher}
\affiliation{Max Planck Institute for the Physics of Complex Systems, N\"othnitzer Stra\ss e 38, 01187, Dresden, Germany}
\affiliation{Cluster of Excellence Physics of Life, TU Dresden, 01062 Dresden, Germany}

\author{Debarghya Banerjee}
\email{banerjee@pks.mpg.de}
\affiliation{Max Planck Institute for the Physics of Complex Systems, N\"othnitzer Stra\ss e 38, 01187, Dresden, Germany}

\date{\today}

\begin{abstract}
Stress-strain constitutive relations in solids with an internal angular degree of freedom can be modelled using Cosserat (also called micropolar) elasticity. In this paper, we explore Cosserat materials that include chiral active components and hence odd elasticity. We calculate static elastic properties and show that the static response to rotational stresses leads to strains that depend on both Cosserat and odd elasticity. We compute the dispersion relations in odd Cosserat materials in the overdamped regime and find the presence of \emph{exceptional points}. These exceptional points create a sharp boundary between a Cosserat-dominated regime of complete wave attenuation and an odd-elasticity-dominated regime of propagating waves. We conclude by showing the effect of Cosserat and odd elastic terms on the polarization of Rayleigh surface waves.
\end{abstract}

\maketitle

The elastic behavior of an isotropic solid at equilibrium can be characterized by two elastic constants, namely the shear modulus and the bulk modulus~\cite{Landauv7}. This simple description of elastic properties using two coefficients is possible because of symmetries such as: isotropy, parity, and time reversal invariance. However, this simple definition does not apply to a variety of other systems, for example nematic solids~\cite{deGennes1993} and Cosserat (or micropolar) solids~\citep{dyszlewicz_micropolar_2004}. Typical elastic solids can be microscopically modelled by considering point masses connected by springs. By contrast, Cosserat elasticity is based on a more complex picture, and includes an angle ($\phi$) describing the microscopic orientational degree of freedom. Even for models consisting of point particles, Cosserat-like elasticity can emerge due to a geometry based on rotating elements~\cite{Sun2012}. Recent advances in additive manufacturing (or 3D printing) have led to rapid developments in the design of metamaterials with Cosserat elasticity~\citep{Rueger2018,Rueger2019,Vasiliev2021,Zhang2018,Giorgio2020,Lakes2017,Lakes2009}. Cosserat elasticity can also emerge in disordered solids ~\citep{Nampoothiri2020,Cates2004,Ganguly2013,Sussman2015,Mitarai2002,Merkel2011}, elastic polymers~\citep{Hiergeist1996,Assidi2011}, and bio-membranes with viscoelastic responses~\citep{Zimmerberg2006,Lakes2001,Salbreux2017}. The Cosserat filament model has been used to explore the effect of microrotations in biological filaments~\citep{Floyd2022,Gunaratne2022,Gazzola2018,Sack2016,Welch2020}.

Active solids~\citep{Maitra2019,shankar2022topological,binysh2022active,brandenbourger2019non,amar2005growth,goriely2017mathematics} are solids that are far from equilibrium due to forcing at the microscopic scales~\cite{Marchetti2013,Ramaswamy2010,Toner2005}. To consider the effect of activity and chirality, a situation that can emerge when activity is present in the form of an active torque, in an elastic Cosserat solid one must include the effect of \emph{odd} elasticity. The odd elasticity is connected to the breaking of two essential symmetries of classical solids -- parity invariance and time reversal invariance -- and appears in the elasticity tensor as a term breaking the major symmetry of the fourth rank elastic tensor, i.e., $\kappa^o_{ijkl} = -\kappa^o_{klij}$. Recent literature~\cite{Banerjee2017,Banerjee2020,Banerjee2020viscoelastic,Souslov2019,Souslov2020, Khain2020,Maitra2020,Kole2020,Markovich2020,Mukhopadhyay2021,Abanov2019,Ganeshan2017,Lier2022, Lin2022, Fodor2021} has extensively studied the effects of odd elasticity and other forms of odd responses in solids and fluids, and it is therefore worthwhile to study the effects of odd elasticity in solids with active torques.

\begin{figure*}
\includegraphics[width=\linewidth]{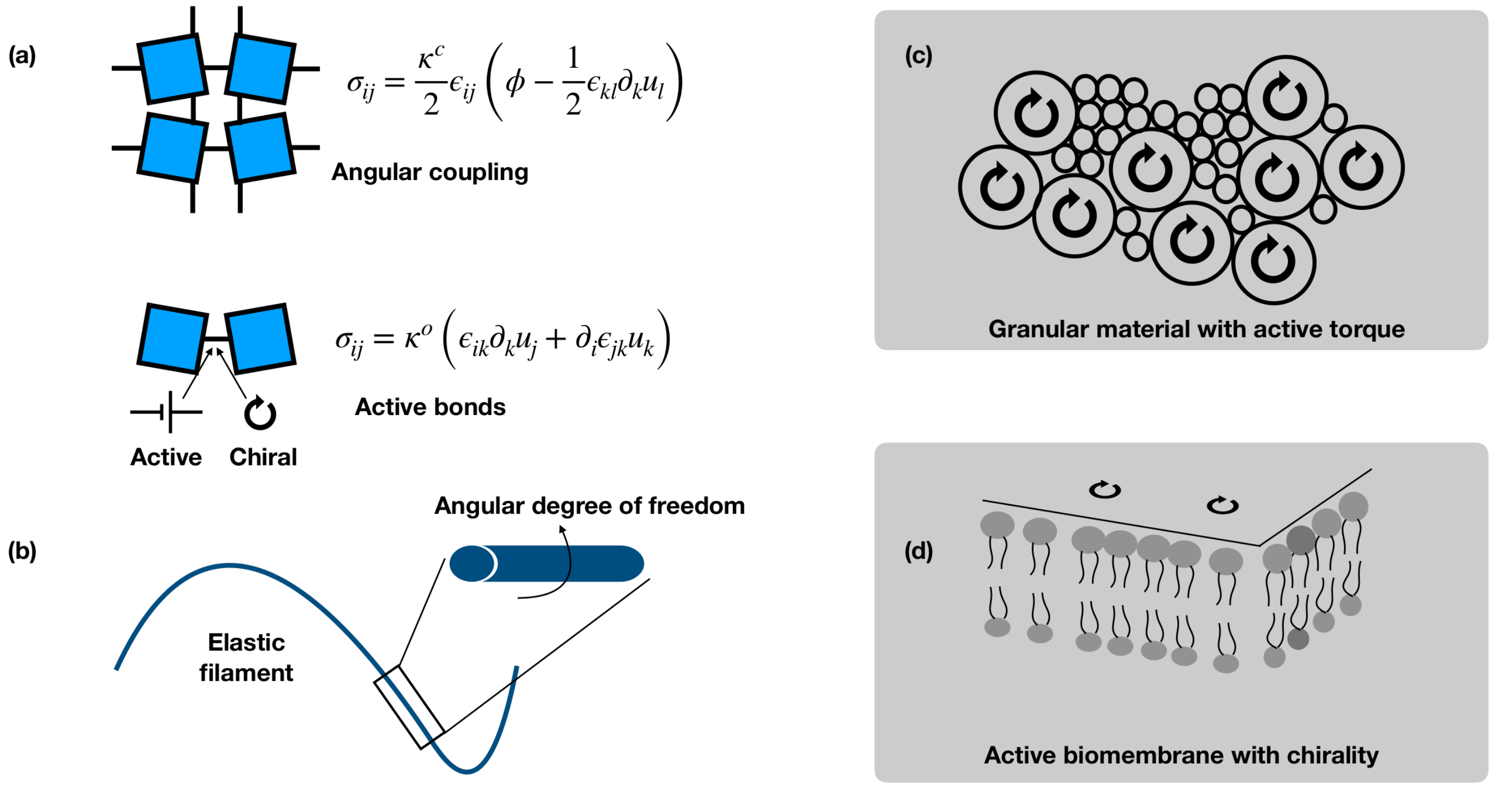}
\caption{ {\bf Odd Cosserat materials}. (a) The presence of extended objects at lattice points connected by springs necessitates an additional angle variable to define the strain. These microrotations are modelled using the so-called Cosserat term. In addition, the bonds can be chiral and active, which can be modelled (to leading order) using so-called odd elasticity. The active bonds illustrated above can be thought of as springs that inject energy and angular momentum, i.e., possess some form of active torque. (b) Rod-like elastic filaments (like actin filaments) have an angular degree of freedom and their elastic properties have been modelled using Cosserat elasticity. (c) Granular material with active torque and (d) active biomembranes with chiral processes are two other examples of odd Coserat materials.}
\label{fig:cosserat_schematic}
\end{figure*}

In this paper, we show how the simultaneous presence of both odd elasticity and the Cosserat term affects the static and dynamic elastic response of chiral active solids in the over-damped regime. We find that the static response to off-diagonal stresses is strongly dependent on both Cosserat and odd elasticity. We also find that dynamic modes have an \emph{exceptional point}~\citep{Bender1998,Heiss2012} in the dispersion relation due to the competition of Cosserat and odd elasticity. In the overdamped regime, this exceptional point is characterised by a transition from damped oscillations to diffusive (or attenuating) solutions. The diffusive solutions near these exceptional points have a diffusion coefficient proportional to the square root of the coefficient of Cosserat elasticity, in contrast to both equilibrium Cosserat solids which have no exceptional points and odd-elastic solids without Cosserat terms. Furthermore, the edge of these solids exhibits edge modes~\citep{Landauv7,Benzoni2021,Rayleigh1885} whose polarization is affected by the combination of odd and Cosserat elasticity. The Cosserat elastic coefficient results in a renormalization of the usual elastic terms for these edge waves, while the odd elastic coefficient mixes the longitudinal and transverse waves at the edge. 
  
{\it Effective theory with odd elasticity} --- We begin by considering the stress tensor in a two-dimensional odd Cosserat solid (Fig.~\ref{fig:cosserat_schematic}). We can write the constitutive relation between stress $\sigma_{ij}$ and strain $u_{ij}$ in these materials as:
\begin{align}
\sigma_{ij} & = \mu u_{ij} + B \delta_{ij} u_{kk} + \frac{\kappa^c}{2} \epsilon_{ij} \left(\phi - \frac{1}{2}\nabla \times {\bf u}\right) \nonumber \\
& +  \kappa^o  \left( \partial_i u_j^* + \partial_i^* u_j \right).
\label{eq:cosserat_stress}
\end{align}
Here, the vector ${\bf u}$ is the displacement field. The stress tensor $\sigma_{ij}$ depends on strain tensor $u_{ij}$, which is defined as the spatial gradients of the displacement $u_{ij} = 1/2 (\partial_i u_j + \partial_j u_i)$, described by the elastic coefficients $\mu$ and $B$.  The coefficient of Cosserat elasticity $\kappa^c$ describes a coupling of the internal displacement gradients to an orientational degree of freedom with angle $\phi$. The two-dimensional Levi-Civita symbol is denoted by $\epsilon_{ij}$. The odd elastic coefficient is denoted by $\kappa^o$ and we define $u_i^*= \epsilon_{ij} u_j$. The odd-Cosserat model described in Eq.\eqref{eq:cosserat_stress} respects rotational invariance.

The dynamics of solids can depend not just on the relation between the elastic stresses and strains, but also on viscous stresses proportional to the strain rates i.e., $\sigma^{vis}_{ij} = \eta_{ijkl} \dot{u}_{kl}$. A combination of viscous and elastic stresses in solids is described by the Kelvin-Voigt model of viscoelasticity, which in turn can be generalised to include odd elastic terms~\cite{Banerjee2020viscoelastic}. However, in this paper we focus on the elastic properties only and hence neglect for simplicity the viscosus stress. In addition to the constitutive relation given in Eq.\eqref{eq:cosserat_stress}, the equation of motion for the displacement field can be written as:
\begin{align}
\rho \partial_t^2 u_i + \Gamma \partial_t u_i &=  \partial_j \sigma_{ij} \nonumber \\
\mathcal{I} \partial_t^2 \phi  + \Gamma_{\phi} \partial_t \phi &= \alpha \nabla^2 \phi - \kappa^c \left( \phi - \frac{1}{2} \nabla \times {\bf u} \right) + \tau^a,
\label{eq:cosserat_EOM}
\end{align}
where $\rho$ is the mass density and $\mathcal{I}$ is the moment of inertia density. The coefficients $\Gamma$ and $\Gamma_{\phi}$ are friction coefficients that arise due to the damping of relative motion with respect to a substrate. The coefficient $\alpha$ is a diffusive coefficient. The term proportional to $\kappa^c$ is required by angular momentum conservation \footnote{We can use $\nabla \times {\bf u}$ and $\nabla^* \cdot {\bf u}$ interchangeably without affecting any physics}. The active torque is denoted by $\tau^a$. The above equations (Eq.~\eqref{eq:cosserat_EOM}) do not take into account non-linear terms.

So far we have discussed the presence of odd elasticity in the equation for $u_{ij}$. We now consider terms that constitute active contributions in the equation of motion for $\phi$. The Cosserat term can be derived from a free energy $ \mathcal{F} = \int d {\bf x} ~ (\kappa^c/2) \left( \phi - (1/2) \nabla \times {\bf u} \right)^2$. Model-A type dynamics~\citep{CL,Hohenberg1977} using this free energy gives us the equations of motion for Cosserat materials (see supplementary section~\cite{Suppli}). In order for a term to qualify as an active contribution, the term has to be such that it cannot be derived from an equilibrium free energy. Therefore, we do not find a linear active term in the equation for $\phi$. The leading order active contribution arising in this equation is nonlinear and given by:
\begin{align}
    \tau^a = \lambda |\nabla \phi|^2 +\ldots.
\end{align}
Mathematically, these terms are similar to the type of active terms that arise in active binary mixtures~\citep{Tjhung2018,Wittkowski2014,Cates2018}~\footnote{The $\phi$ here is a pseudo-scalar while the concentration field in the binary mixture problem is a scalar field this implies that the coefficient $\lambda$ is a pseudo-scalar}. An important point to be noted here is that because of the positive definite nature of the term, the sign of $\lambda$ decides the sense of the active torque, making the system naturally chiral. However, in the presence of either polar or nematic activity, it is possible to have linear active terms. It is also possible to model active torques in the form of torque dipoles~\citep{Furthauer2012,Furthauer2013,Naganathgan2014,Markovich2019}.

{\it Elastostatics of odd Cosserat materials} --- Using the stress tensor in Eq.\eqref{eq:cosserat_stress}, we now discuss the static properties of odd Cosserat materials. Since we are considering the static regime, all time derivatives drop out. In static equilibrium, solids balance external stresses by the elastic stresses, which are proportional to the strain. In the case of Cosserat solids, there is an additional degree of freedom, which is the angle $\phi$. This force balance can be written as a set of linear equations with the applied stress on one side and the internal stress on the other side. If we now additionally neglect the higher order spatial gradients arising due to diffusion we then have a linear problem where we can choose a profile of external stress and from that obtain the strain in static equilibrium. Material properties like the Young's modulus ($E$), Poisson ratio ($\nu$), and odd ratio ($\nu^o$, defining the transverse tilt of the solid under uniaxial compression, see Ref.~\cite{Scheibner2020}) can be computed using this method. We find the emergence of auxetic properties (negative value of $\nu$) in the limit of large odd elasticity ($2 (\kappa^o)^2 > \mu B$), which has been previously reported for non-Cosserat odd elastic solids in Ref.~\citep{Scheibner2020}. Details of this computation is provided in ~\cite{Suppli}.

While the moduli ($E$, $\nu$, and $\nu^o$) remain largely unaffected under the application of a uniaxial pressure, we can obtain generic expressions of strain in the presence of applied stress. Let us consider a problem where we have only applied rotational and transverse stresses, i.e., only $\sigma_{xy}$ and $\sigma_{yx}$ are non-zero. Under such an external stress, components of the strain tensor have the form:
\begin{align}
    & \frac{\partial u_x}{\partial x} = -\frac{\kappa^o (v_1 + v_2)}{4 {\kappa^o}^2 + \mu^2}, \nonumber \\
    & \frac{\partial u_y}{\partial y} = \frac{\kappa^o (v_1 + v_2)}{4 {\kappa^o}^2 + \mu^2}, \nonumber \\
    & \frac{\partial u_y}{\partial x} = \frac{v_2 - v_1}{\varepsilon} + \frac{v_2 - v_1}{\kappa^c} + \frac{\mu (v_1 + v_2)}{2 (4 {\kappa^o}^2 + \mu^2)}, \nonumber \\
    & \frac{\partial u_x}{\partial y} = -\frac{v_2 - v_1}{\varepsilon} - \frac{v_2 - v_1}{\kappa^c} + \frac{\mu (v_1 + v_2)}{2 (4 {\kappa^o}^2 + \mu^2)},
\end{align}
where, $v_1 = \sigma_{xy}$, $v_2 = \sigma_{yx}$, and $\varepsilon$ is a regularizing term added to the equation of motion of $\phi$ to make the matrix invertible. Physically, one can think of $\varepsilon$ as an effective renormalization of the higher order spatial gradients that have been neglected and also from coupling to external substrate.

{\it Elastodynamics of odd Cosserat materials} --- We now consider the dynamics of the system described by Eq.~\eqref{eq:cosserat_stress} and Eq.~\eqref{eq:cosserat_EOM}. In the under-damped limit (i.e., $\Gamma \rightarrow 0$, $\Gamma_{\phi} \rightarrow 0$, and ignoring other viscous effects), we can obtain oscillatory solutions from the elastic terms. We now consider propagating waves of the form $\exp(i({\bf k} \cdot {\bf x} - \omega t ))$, where ${\bf x}$ is the position coordinate and ${\bf k}$ is the wavevector which gives us wavenumber $k = |{\bf k}|$. The frequency is given by $\omega$ and $t$ is time. In the limit of large odd elasticity, we obtain a dispersion relation: $\omega = e^{i \pi/4} k \sqrt{\kappa^o/\rho} $, where, $\omega$ is the frequency and $k$ is the wavenumber. The above dispersion relation corresponds to a propagating wave in displacement with speed $\sqrt{\kappa^o/2 \rho}$ and a damping constant $\sqrt{\kappa^o/2 \rho}$. In the case of a normal elastic solid without any Cosserat coupling or odd elasticity, we have the usual elastic waves with dispersion relations $\omega = k \sqrt{\mu/\rho}$ and $\omega = k \sqrt{(B +\mu)/\rho}$. Therefore, unlike normal elastic waves, in odd elastic solids, we have spontaneous damping and injection of energy in the form of linear displacement fluctuations within the solid. 

We now consider the over-damped regime, where we can neglect inertial effects and absorb the damping coefficients ($\Gamma$ and $\Gamma_{\phi}$) into the other coefficients of the problem. In typical elastic materials, such an over-damped limit gives rise to damped modes due to the effects of the shear modulus and bulk modulus. Odd elasticity gives rise to propagating waves, which are analogous to the \emph{Avron waves} in odd-viscous fluids~\citep{Avron1998,Lucas2014,Ganeshan2017}. However, for an odd solid these are waves in displacement and not in velocity.  We now compute the dispersion relation for a Cosserat solid in the presence of odd elasticity. For the purpose of this calculation, we ignore the bulk modulus and consider the limit where we can neglect the fluctuations in $\phi$, i.e., $\phi=\phi_0$. We obtain a closed form expression of the dispersion relation given by:
\begin{align}
    \omega = i \frac{k^2}{8} \left(-8 \mu - \kappa^c \pm \sqrt{{\kappa^c }^2 - 64 {\kappa^o}^2}\right),
\end{align}
which is consistent with our understanding that the Cosserat term is predominantly a diffusive term in the equations of elastodynamics, whereas odd elasticity gives rise to wave solutions with speed proportional to $\sqrt{\kappa^o}$. Note that we obtain an exceptional point at $\kappa^c = 8 \kappa^o$, where the eigenvalues are degenerate and have a square root branch point~\citep{Bender1998,Heiss2012}. At the exceptional point, two eigenvectors coalesce to a single one and the eigenvalues are degenerate. At this exceptional point, we find the transition from a diffusive solution with diffusivity proportional to $\sqrt{\kappa^c}$ to a damped wave solution with speed proportional to $\sqrt{\kappa^o}$. This is the key differentiating feature of the odd Cosserat elasticity from both equilibrium elastic solids and active odd elastic solids.

\begin{figure}[htbp!]
    \centering
    \includegraphics[width=\linewidth]{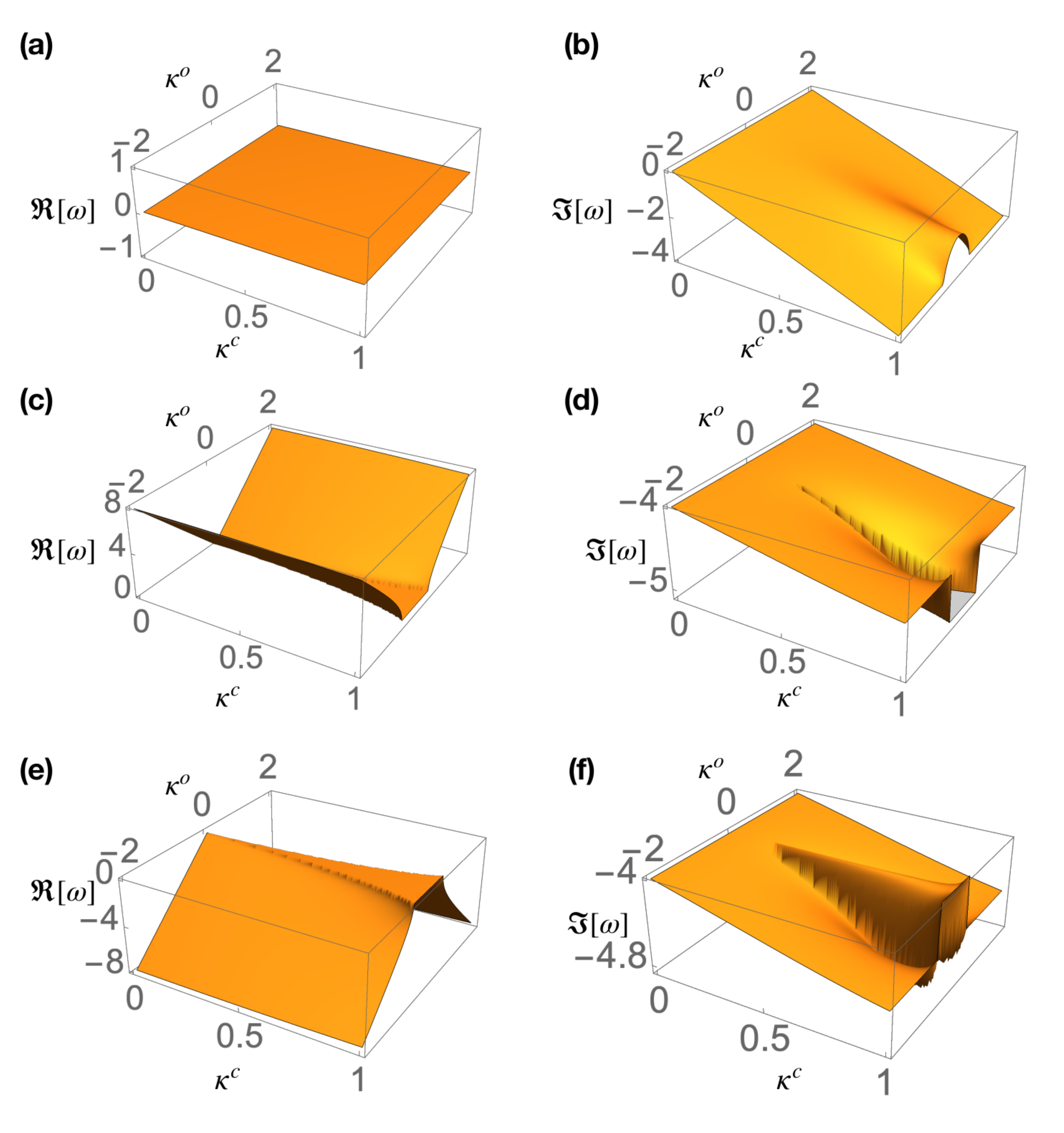}
    \caption{{\bf Dependence of frequency on odd and Cosserat elasticity}. Real and imaginary parts of the dispersion relation for an odd Cosserat solid. The three solutions of $\omega$ (three eigenfrequencies) have both real and imaginary parts. The first row, i.e., (a) and (b), is the eigenfrequency corresponding to the first eigenmode, the second row, i.e., (c) and (d), is the eigenfrequency corresponding to the second eigenmode, and the third row, i.e., (e) and (f), is the eigenfrequency corresponding to the third eigenmode. The first column (a),(c), and (e) shows the behavior of the real part and the second column (b),(d), and (f) shows the behavior of the imaginary part. The real part corresponds to oscillations and emerges only for finite odd elasticity in both the second and the third eigenmodes. For these plots, we have fixed the values of $\mu$ and $k$ at $\mu=1$ and $k=1$.}
    \label{fig:cosserat_dispersion}
\end{figure}

For the usual Cosserat solids with $\kappa^o = 0$, the exceptional points are absent because of the Hermitian nature of the matrix. The discriminant of the cubic equation does not take negative values because it can be written as a sum of positive numbers. However, in the more general case, if we take into account the effect of both $\kappa^c$ and $\kappa^o$ we obtain $\omega = i \Lambda$ that are solutions to the cubic equation (see supplementary material~\cite{Suppli} for calculation):
\begin{align}
    & \Lambda^3 +  \left( 4 \kappa^c + \kappa^c k^2 + 8 \mu k^2 \right) \Lambda^2 \nonumber \\
    & + \left( 32 \mu \kappa^c k^2 + 4 (4 \mu^2 + \mu \kappa^c + 4 {\kappa^o}^2) k^4 \right) \Lambda \nonumber \\
    & + 64 \kappa^c ({\kappa^o}^2 + \mu^2) k^4 = 0.
\end{align}
In the supplementary material~\cite{Suppli}, we derive the condition for a positive discriminant based on the solutions to this cubic equation. In Fig.~\ref{fig:cosserat_dispersion}, we show the real and imaginary part of the dispersion relations showing the emergence of a non-zero imaginary part for finite odd elasticity as a signature of the exceptional points.

{\it Rayleigh waves} --- We consider the effect of both the Cosserat and the odd terms on a stress-free edge. We consider solutions of the form ${\bf u} = {\bf U} e^{i(kx - \omega t)} e^{a y}$ (and ignore fluctuations in $\phi$) and a boundary on the line parallel to the $x$-axis, i.e., the $y$-direction is normal to the edge. Implementing a zero normal stress implies $\sigma_{yy} = \sigma_{xy} = 0$ (and $\phi = 0$). We find that while Cosserat elasticity renormalizes parameters, the polarization of these edge waves depends crucially on the presence and nature of odd elasticity. We perform the detailed computation in Ref.~\cite{Suppli}.

The difficulty in solving problems pertaining to the edge is that the edges are naturally asymmetric, i.e., the direction along the edge and perpendicular to the edge are qualitatively different. We now define two types of waves: one which is transverse and one which is longitudinal. The decay length along the $y$-direction for the transverse and longitudinal waves are $a_t$ and $a_l$, respectively. The amplitudes are given by $U_t$ and $U_l$, and we obtain the following relations:
\begin{align}
   & U_t \left[ -i 2k \kappa^o a_t + \frac{\mu}{2} (k^2 + a_t^2) -\frac{\kappa^c}{4} (k^2 - a_t^2) \right] + \nonumber \\
   & U_l \left[ -i \kappa^o (k^2 + a_t^2)  + \mu k a_l \right] = 0, \nonumber \\
   & U_t \left[  \mu k a_t - i \kappa^o (k^2 + a_t^2) \right] + \nonumber \\ 
   & U_l \left[ - B (k^2 - a_l^2) + \mu a_l^2 - i 2 \kappa^o k a_l\right] = 0.
\end{align}
In the above relations, we find that the odd elasticity makes the equations complex. The amplitudes acquire phases proportional to the odd elasticity, thus indicating that the Rayleigh surface waves acquire a phase proportional to $\kappa^o$. Physically, this occurs because the odd elasticity term mixes the longitudinal and the transverse waves at the edge while the Cosserat term has a much less drastic effect of simply renormalising the equilibrium elastic constants. For a chiral Cosserat solid without odd elasticity, the chiral effects can be incorporated in the solutions by choosing a different boundary condition for $\phi$ (instead of $\phi=0$ as we have done above). Therefore, in an odd Cosserat solid, we obtain surface waves that are in a mixed state of the usual longitudinal and transverse waves and the effective parameters are renormalized due to the presence of the Cosserat elasticity.  

{\it Conclusions} --- To conclude, we have studied in this paper the effects of adding odd elasticity to Cosserat solids. We find while studying the static response that strain at mechanical equilibrium acquires components that, generically, are dependent on both odd and Cosserat elasticity. The linear dispersion relations of odd Cosserat materials exhibit exceptional points where the solutions change from diffusive to propagating waves. Previous studies of odd elastic solids have also observed exceptional points, but in odd Cosserat materials, the exceptional points arise due to a competition between odd elasticity and Cosserat elasticity. A signature of these exceptional points appears in the surface waves by the generation of waves which are mixtures of transverse and longitudinal modes. We envision experimental verification of our results in robotic metamaterials, disordered solids, and active gels.

\textit{Acknowledgements} --- DB would like to thank Subhro Bhattacharjee,  Bulbul Chakraborty, Ruben Lier, Ananyo Maitra, Colin Scheibner,  Peter Sollich, Suropriya Saha, Dario Vincenzi, Vincenzo Vitelli, and Piotr Witkowski for illuminating discussions.  PS is supported in part by the Narodowe Centrum Nauki (NCN) Sonata Bis grant 2019/34/E/ST3/00405 and the Deutsche Forschungsgemeinschaft (DFG) through the cluster of excellence ct.qmat (Exzellenzcluster 2147, Project 390858490). A.S.~acknowledges the support of the Engineering and Physical Sciences Research Council (EPSRC) through New Investigator Award No.~EP/T000961/1.

\begin{widetext}

\section{Cosserat and angular momentum conservation (passive solids)}

The under-damped equation of motion in two dimensional space(neglecting other terms like viscosity e.t.c.) in the presence of Cosserat stress is given by:
\begin{align}
\rho \partial_t^2 u_i &=  \partial_j \sigma_{ij} \nonumber \\
\mathcal{I} \partial_t^2 \phi &= \alpha \nabla^2 \phi - \kappa^c \left( \phi - \frac{1}{2} \nabla \times {\bf u} \right),
\label{eqA:cosserat_EOM}
\end{align}
where,
\begin{align}
\sigma^c_{ij} & = \frac{\kappa^c}{2} \epsilon_{ij} \left( \phi - \frac{1}{2}\nabla \times {\bf u} \right),
\label{eqA:cosserat_stress}
\end{align}
the above equations conserve angular momentum in spite of the presence of an anti-symmetric stress (shown in Eq.~\ref{eqA:cosserat_stress}) because of the
term with coefficient $\kappa^c$ in the equations of motion for local angular momentum.If one considers a closed loop integral of the above stress one can show that the net torque injected due to this stress and is given by $\sigma_{xy}-\sigma_{yx}$ which is compensated by the $\kappa^c$ term in the evolution equation for local angular momentum. This argument also works for micropolar fluids where the $\phi$ is replaced by the rate of change of $\phi$ (we can call it $\Omega$) the displacement vector is replaced by the velocity vector.

In over-damped case the above equations reach a state where the local angle relaxes to the curl of the strain i.e. $2 \phi = \nabla \times {\bf u}$ and one reaches a state where the anti-symmetric stress vanishes and the local-angular momentum evolution equation becomes irrelevant.  For problems relating to micropolar fluids similar arguments can be made and twice the local angular momentum relaxes to vorticity of the fluid in finite time.  

Thus, we find that even for passive solids the under-damped equation may have important contribution from anti-symmetric stress in the equation of motion.  

\section{Is symmetrization of stress tensor possible in passive Cosserat solids? }

Linear momentum due to the strain rate is defined as $g_i = \rho \dot{u}_i$ and hence the conservation law is given by $\rho \ddot{u}_i = \partial_j \sigma_{ij}$. For simplicity we consider a system with only the stress given in Eq.~\ref{eqA:cosserat_stress}. Let us consider the following transformations:
\begin{align}
& g_i' = g_i + \frac{1}{2} \partial_j \epsilon_{ij} \mathcal{I} \dot{\phi}, \nonumber \\
& \phi' = 0 , \nonumber \\
& \sigma_{ij}' = 0
\end{align}
In fluid systems this simply gives us the equations of motion in the primed variables. However, in solids this argument may not be simply used because the under-damped motion in solids do not give relaxation dynamics and hence anti-symmetric modes remain intact and the over-damped equations are not conservation laws.

\section{Equilibrium elastic theory with an angle}

Let us begin by first deriving the Gibbs-Duhem relations and equilibrium stress for a Cosserat viscoelastic material. The free energy can be written as:
\begin{align}
    F = \int d {\bf x} \left[ \frac{1}{2} \rho \dot{u}_i \dot{u}_i + \frac{1}{2} I \rho \dot{\phi}^2 + f_0(\partial_i u_j, \partial_i \phi,\phi) \right],
\end{align}
in the above free energy we have used $\mathcal{I} \equiv I \rho$, where $I$ is a proportionality coefficient and this assumption states that moment of inertia density is proportional to mass density. We can also define chemical potential like terms as:
\begin{align}
    & \mu_{\phi} \equiv \frac{\partial f_0}{\partial \phi} - \partial_i \frac{\partial f_0}{\partial (\partial_i \phi)}, \nonumber \\
    & \mu_j \equiv -\partial_i \frac{\partial f_0}{\partial (\partial_i u_j)}
\end{align}
Under infinitesimal spatial translations the change in free energy is:
\begin{align}
    \delta F = \int d {\bf x} \left( \rho \dot{u}_i \delta \dot{u}_i + \frac{\dot{u}_i \dot{u}_i}{2} \delta \rho + I \rho \dot{\phi} \delta \dot{\phi}  + \frac{I \dot{\phi}^2}{2} \delta \rho + \mu_{\phi} \delta \phi + \mu_i \delta u_i\right) + \oint d s_{\alpha} \left( f \delta x_{\alpha} + \frac{\partial f_0}{\partial (\partial_{\alpha} \phi)} \delta \phi + \frac{\partial f_0}{\partial (\partial_{\alpha} u_i)} \delta u_i \right)
\end{align}
where, $f \equiv \frac{1}{2} \rho \dot{u}_i \dot{u}_i + \frac{1}{2} I \rho \dot{\phi}^2 + f_0(\partial_i u_j, \partial_i \phi,\phi)$. Now, we use the following relations for infinitesimal transformations:
\begin{align}
    & \delta \rho = - \delta x_{\alpha} \partial_{\alpha} \rho \nonumber \\
    & \delta \dot{u}_i = -\delta x_{\alpha} \partial_{\alpha} \dot{u}_i \nonumber \\
    & \delta \dot{\phi} = -\delta x_{\alpha} \partial_{\alpha} \dot{\phi} \nonumber \\
    & \delta \phi = -\delta x_{\alpha} \partial_{\alpha} \phi \nonumber \\
    & \delta u_i = -\delta x_{\alpha} \partial_{\alpha} u_i
\end{align}
Using the above expressions and by using divergence theorem we obtain:
\begin{align}
    \delta F = \int d {\bf x} \left( (\partial_{\beta} \mu_{\phi}) \phi + (\partial_{\beta} \mu_i ) u_i \right) \delta x_{\beta} + \oint d s_{\alpha} \left( (f_0 - \phi \mu_{\phi} - \mu_i u_i) \delta_{\alpha \beta} \delta x_{\beta} - \frac{\partial f_0}{\partial (\partial_{\alpha} \phi)} (\partial_{\beta} \phi) \delta x_{\beta}  - \frac{\partial f_0}{\partial (\partial_{\alpha} u_i)} (\partial_{\beta} u_i) \delta x_{\beta} \right)
\end{align}
From the above expression for free energy change in a infinitesimal translation we obtain the equilibrium stress as below:
\begin{align}
    \sigma^e_{\alpha \beta} = (f_0 - \phi \mu_{\phi} - \mu_i u_i) \delta_{\alpha \beta} - \frac{\partial f_0}{\partial (\partial_{\alpha} \phi)} (\partial_{\beta} \phi)  - \frac{\partial f_0}{\partial (\partial_{\alpha} u_i)} (\partial_{\beta} u_i)
\end{align}
and the Gibbs-Duhem relation:
\begin{align}
    -\partial_{\beta} \sigma^e_{\alpha \beta} = (\partial_{\alpha} \mu_{\phi}) \phi + (\partial_{\alpha} \mu_i ) u_i 
\end{align}

\subsection{Irreversible thermodynamics and Cosserat viscoelasticity}

For a non equilibrium system the entropy evolution is given by:
\begin{align}
    \partial_t s + \partial_{\alpha} J_{\alpha}^{s} = \theta,
\end{align}
where, $s$ is the entropy density, $J_{\alpha}^{s} = s v_{\alpha} + j_{\alpha}^s$ is the entropy flux, $j_{\alpha}^s$ is the relative entropy flux in the COM frame, and $\theta > 0$ is the entropy production rate per unit volume due to irreversible processes. Similarly the free energy density follows:
\begin{align}
    \partial_t f + \partial_{\alpha} J_{\alpha}^f = \theta_f,
\end{align}
where, $J_{\alpha}^f$ is the free energy flux and $\theta_f$ is the source of free energy. The free energy density obeys the local thermodynamic relation $f = e - T s$. The relative flux in the COM frame is $j_{\alpha}^f$ and we have the relation $J_{\alpha}^f = f v_{\alpha} + j_{\alpha}^f$. The total energy flux is the sum of free energy transport and heat transport i.e. $J_{\alpha}^e = (f + Ts) v_{\alpha} + j_{\alpha}^f + j_{\alpha}^Q$. We define $j_{\alpha}^Q = T j_{\alpha}^s$. Thus free energy flux is $j_{\alpha}^f = j_{\alpha}^e - j_{\alpha}^Q$ is part of the relative energy flux that is nor heat. In isothermal system at temperature $T$ the local reduction of free energy is directly related to entropy production: $T \theta = -\theta_f$. 

Let us now derive the constitutive relations of Cosserat viscoelasticity from the principles of irreversible thermodynamics. To begin with a free energy given by:
\begin{align}
    F = \int d {\bf x} \left[ \frac{1}{2} \rho \dot{u}_i \dot{u}_i + \frac{1}{2} I \rho \dot{\phi}^2 + f_0(\partial_i u_j, \partial_i \phi,\phi) \right] \nonumber \\
    f_0 \equiv \frac{\kappa^c}{2} \left(\phi - \frac{1}{2} \epsilon_{kl} \partial_k u_l \right)^2 + \kappa_{ijkl} u_{kl} u_{ij} + \frac{\alpha}{2} (\partial_i \phi) (\partial_i \phi)
\end{align}
Therefore, we have the following:
\begin{align}
    &\frac{\partial f_0}{\partial \phi} = \kappa^c \left( \phi - \frac{1}{2} \epsilon_{kl} \partial_k u_l \right), \nonumber \\
    &-\partial_i \frac{\partial f_0}{\partial (\partial_i \phi)} = - \alpha \partial_i \partial_i \phi, \nonumber \\
    &- \partial_i \frac{\partial f_0}{\partial (\partial_i u_j)} = \frac{\kappa^c}{2} \epsilon_{ij} \partial_i \left( \phi - \frac{1}{2} \epsilon_{kl} \partial_k u_l \right) -\partial_i \left( \kappa_{ijkl} \frac{1}{2}(\partial_k u_l + \partial_l u_k ) \right),
\end{align}
and the conservation laws given by:
\begin{align}
    &\partial_t (\rho \dot{u}_i) = - \partial_j \sigma_{ij}, \nonumber \\
    &\partial_t \rho = - \partial_j (\rho \dot{u}_j), \nonumber \\
    &\partial_t (I \rho \dot{\phi}) = \epsilon_{ij} \sigma_{ij} - \partial_j \chi_j.
\end{align}

The free energy evolution is given by:
\begin{align}
    \dot{F} = & \int d {\bf x} \left[ \frac{\dot{u}_i^2}{2} \dot{\rho} + \dot{u}_i \partial_t (\rho \dot{u}_i) + \frac{I}{2} \dot{\phi}^2 \dot{\rho}  + \dot{\phi} \partial_t (I \rho \dot{\phi}) + \dot{\phi} \left( \frac{\partial f_0}{\partial \phi} - \partial_i \frac{\partial f_0}{\partial (\partial_i \phi)}\right) - \dot{u}_j \left( \partial_i \frac{\partial f_0}{\partial (\partial_i u_j)}\right)\right] \nonumber \\
    & \int d {\bf x} \left[ \left( -\frac{\dot{u}_i \dot{u}_i}{2} - \frac{I \dot{\phi}^2}{2}\right) \partial_j (\rho \dot{u}_j) - \dot{u}_i \left( \partial_j \sigma_{ij} + \frac{\kappa^c}{2} \epsilon_{ij} \partial_j \left( \phi - \frac{1}{2} \epsilon_{kl} \partial_k u_l\right) - \partial_j \left( \kappa_{ijkl} \frac{1}{2}(\partial_k u_l + \partial_l u_k ) \right)\right) \right. \nonumber \\ 
    &+ \left. \dot{\phi} \left( \epsilon_{ij} \sigma_{ij} - \partial_j \chi_j + \kappa^c \left( \phi - \frac{1}{2} \epsilon_{kl} \partial_k u_l\right) - \alpha \partial_k \partial_k \phi\right) \right] 
\end{align}

If we now consider the integrand and look at it part by part then the first part is given by:

\begin{align}
    &-\frac{1}{2} (\dot{u}_i \dot{u}_i + I \dot{\phi} \dot{\phi}) \partial_j (\rho \dot{u}_j) \nonumber \\
    &= -\frac{1}{2} \partial_j \left[ \dot{u}_i \dot{u}_i \rho \dot{u}_j + I \dot{\phi}^2 \rho \dot{u}_j\right] + \frac{1}{2} \rho \dot{u}_j \partial_j (\dot{u}_i \dot{u}_i + I \dot{\phi}^2)
\end{align}
This gives us the integral:
\begin{align}
    \int d {\bf x} \left( \rho \dot{u}_i \dot{u}_j \partial_j \dot{u}_i + I \rho \dot{\phi} \dot{u}_j \partial_j \dot{\phi}\right) + {\rm Surface~terms}
\end{align}

The second part is given by:

\begin{align}
    &- \dot{u}_i \left( \partial_j \sigma_{ij} + \frac{\kappa^c}{2} \epsilon_{ij} \partial_j \left( \phi - \frac{1}{2} \epsilon_{kl} \partial_k u_l\right) - \partial_j \left( \kappa_{ijkl} \frac{1}{2}(\partial_k u_l + \partial_l u_k ) \right)  \right) \nonumber \\
    &= -\partial_j \left( \dot{u}_i \left( \sigma_{ij} + \frac{\kappa^c}{2} \epsilon_{ij} \left( \phi - \frac{1}{2} \epsilon_{kl} \partial_k u_l\right)       - \left( \kappa_{ijkl} \frac{1}{2}(\partial_k u_l + \partial_l u_k ) \right) \right)\right) \nonumber \\
    & + \partial_j \dot{u}_i \left( \sigma_{ij} + \frac{\kappa^c}{2} \epsilon_{ij} \left( \phi - \frac{1}{2} \epsilon_{kl} \partial_k u_l\right) -  \left( \kappa_{ijkl} \frac{1}{2}(\partial_k u_l + \partial_l u_k ) \right) \right)
\end{align}

Which gives the integral:

\begin{align}
    \int d {\bf x}   \partial_j \dot{u}_i \left( \sigma_{ij} + \frac{\kappa^c}{2} \epsilon_{ij} \left( \phi - \frac{1}{2} \partial_{kl} \partial_k u_l\right) - \left( \kappa_{ijkl} \frac{1}{2}(\partial_k u_l + \partial_l u_k ) \right) \right) + {\rm Surface~terms}
\end{align}

Finally, the third part is:
\begin{align}
    & \dot{\phi} \left( \epsilon_{ij} \sigma_{ij} - \partial_j \chi_j + \kappa^c \left( \phi - \frac{1}{2} \epsilon_{kl} \partial_k u_l\right) - \alpha \partial_k \partial_k \phi\right) \nonumber \\
    & \dot{\phi} \left( \epsilon_{ij} \sigma_{ij} + \kappa^c \left( \phi - \frac{1}{2} \epsilon_{kl} \partial_k u_l\right) \right) - \partial_k \left( \dot{\phi} \left( \chi_k + \alpha \partial_k \phi\right)\right) + \left( \chi_k + \alpha \partial_k \phi\right) \partial_k \dot{\phi}
\end{align}
This gives the integral:
\begin{align}
    \int d {\bf x} \dot{\phi} \left( \epsilon_{ij} \sigma_{ij} + \kappa^c \left( \phi - \frac{1}{2} \epsilon_{kl} \partial_k u_l\right) \right) + \left( \chi_k + \alpha \partial_k \phi\right) \partial_k \dot{\phi} + {\rm Surface~terms}
\end{align}

Therefore, we get the final form of the integral excluding the surface terms:

\begin{align}
    & \dot{F} = \int d {\bf x} ~\partial_j \dot{u}_i \left[ \rho \dot{u}_i \dot{u}_j + \sigma_{ij} + \frac{\kappa^c}{2} \epsilon_{ij} \left( \phi - \frac{1}{2} \epsilon_{kl} \partial_k u_l \right) - \left( \kappa_{ijkl} \frac{1}{2}(\partial_k u_l + \partial_l u_k ) \right) \right] + \partial_j \dot{\phi} \left[ I \rho \dot{\phi} \dot{u}_j + \chi_j + \alpha \partial_j \phi\right]  \nonumber \\
    & + \dot{\phi} \left[ \epsilon_{ij} \sigma_{ij} + \kappa^c \left( \phi - \frac{1}{2} \epsilon_{kl} \partial_k u_l\right) \right] 
\end{align}

\begin{align}
{\rm Flux} &\leftrightarrow {\rm Force} \nonumber \\
 \rho \dot{u}_i \dot{u}_j + \sigma_{ij} + \frac{\kappa^c}{2} \epsilon_{ij} \left( \phi - \frac{1}{2} \epsilon_{kl} \partial_k u_l \right) - \left( \kappa_{ijkl} \frac{1}{2}(\partial_k u_l + \partial_l u_k ) \right) & \leftrightarrow  \partial_j \dot{u}_i\nonumber \\
 I \rho \dot{\phi} \dot{u}_j + \chi_j + \alpha \partial_j \phi & \leftrightarrow \partial_j \dot{\phi} \nonumber \\
 \epsilon_{ij} \sigma_{ij} + \kappa^c \left( \phi - \frac{1}{2} \epsilon_{kl} \partial_k u_l\right) &\leftrightarrow \dot{\phi}.
\end{align}

Therefore, we obtain the phenomenological equations:
\begin{align}
    & \rho \dot{u}_i \dot{u}_j + \sigma_{ij} + \frac{\kappa^c}{2} \epsilon_{ij} \left( \phi - \frac{1}{2} \epsilon_{kl} \partial_k u_l \right) - \left( \kappa_{ijkl} \frac{1}{2}(\partial_k u_l + \partial_l u_k ) \right) = \eta \partial_j \dot{u}_i + \lambda_1 \partial_i \partial_j \dot{\phi},  \nonumber \\
    & I \rho \dot{\phi} \dot{u}_j + \chi_j + \alpha \partial_j \phi = \lambda_2 \partial_i \partial_j \dot{u}_i + \lambda_3 \partial_j \dot{\phi}, \nonumber \\
    & \epsilon_{ij} \sigma_{ij} + \kappa^c \left( \phi - \frac{1}{2} \epsilon_{kl} \partial_k u_l\right) = \lambda_4 \dot{\phi} + \lambda_5 \partial_j \partial_j \dot{\phi} + \lambda_6 \partial_i \partial_j \partial_j \dot{u}_i.
\end{align}

If we now consider the part of force and flux that are even under time reversal symmetry we get :
\begin{align}
    & \rho \dot{u}_i \dot{u}_j + \sigma^{\rm reactive}_{ij} + \frac{\kappa^c}{2} \epsilon_{ij} \left( \phi - \frac{1}{2} \epsilon_{kl} \partial_k u_l \right) - \left( \kappa_{ijkl} \frac{1}{2}(\partial_k u_l + \partial_l u_k ) \right) = 0,  \nonumber \\
    & I \rho \dot{\phi} \dot{u}_j + \chi^{\rm reactive}_j + \alpha \partial_j \phi = 0, \nonumber \\
    & \epsilon_{ij} \sigma^{\rm reactive}_{ij} + \kappa^c \left( \phi - \frac{1}{2} \epsilon_{kl} \partial_k u_l\right) = 0.
\end{align}
Similarly, if we consider the part of flux and force that is odd under time reversal symmetry we get:
\begin{align}
    & \sigma^{\rm dissipative}_{ij}  = \eta \partial_j \dot{u}_i + \lambda_1 \partial_i \partial_j \dot{\phi},  \nonumber \\
    & \chi^{\rm dissipative}_j = \lambda_2 \partial_i \partial_j \dot{u}_i + \lambda_3 \partial_j \dot{\phi}, \nonumber \\
    & \epsilon_{ij} \sigma^{\rm dissipative}_{ij}  = \lambda_4 \dot{\phi} + \lambda_5 \partial_j \partial_j \dot{\phi} + \lambda_6 \partial_i \partial_j \partial_j \dot{u}_i.
\end{align}

In the equilibrium limit of a solid all explicit time derivatives are zero and we obtain the equilibrium limit:
\begin{align}
    & \sigma_{ij} + \frac{\kappa^c}{2} \epsilon_{ij} \left( \phi - \frac{1}{2} \epsilon_{kl} \partial_k u_l \right) - \left( \kappa_{ijkl} \frac{1}{2}(\partial_k u_l + \partial_l u_k ) \right) = 0,  \nonumber \\
    & \chi_j + \alpha \partial_j \phi = 0, \nonumber \\
    & \epsilon_{ij} \sigma_{ij} + \kappa^c \left( \phi - \frac{1}{2} \epsilon_{kl} \partial_k u_l\right) = 0.
\end{align}
Thus, we obtain the equilibrium theory of Cosserat elastic solids. I have assumed that all the coefficients are even under time reversal in the above discussion. Another thing to point out here is that the theory is very similar to the Kelvin-Voigt type theory of viscoelastic solids.

\section{Elastostatics calculation}

The linear problem of elastostatics can be defined by the equations below:
\begin{align}
\begin{pmatrix}
B+\mu & B & \kappa^o & \kappa^o & 0\\
B & B+\mu & -\kappa^o & -\kappa^o & 0\\
-\kappa^o & \kappa^o & \frac{\mu}{2} - \frac{\kappa^c}{4} & \frac{\mu}{2} + \frac{\kappa^c}{4} & \frac{\kappa^c}{2} \\
-\kappa^o & \kappa^o & \frac{\mu}{2} + \frac{\kappa^c}{4} & \frac{\mu}{2} - \frac{\kappa^c}{4} & -\frac{\kappa^c}{2} \\
0 & 0 & \frac{\kappa^c}{2} & -\frac{\kappa^c}{2} & -\kappa^c - \varepsilon
\end{pmatrix}
\begin{pmatrix}
\partial_x u_x \\
\partial_y u_y \\
\partial_x u_y \\
\partial_y u_x \\
\phi
\end{pmatrix}
= \begin{pmatrix}
p_1 \\
p_2 \\
v_1 \\
v_2 \\
\theta
\end{pmatrix}
\end{align}
from the above description we can extract the Young's modulus ($E$), Poisson ratio ($\nu$), and odd ratio ($\nu^o$) by setting $p_1=v_1=v_2=\theta = 0$ and $p_2 = p$ (Uniaxial stress along $y$-direction). We obtain:
\begin{align}
 E \equiv \frac{p}{\partial_y u_y} &= \frac{( 4 {\kappa^o}^2  + \mu^2)(\mu + 2 B )}{2 {\kappa^o}^2 + \mu^2  + \mu B}, \nonumber \\
 \nu \equiv -\frac{\partial_x u_x}{\partial_y u_y} &= \frac{-2 {\kappa^o}^2 + \mu B }{2 {\kappa^o}^2 + \mu^2 + \mu B}, \nonumber \\
 \nu^o \equiv -\frac{\partial_y u_x}{2 \partial_y u_y} &=
 \frac{\kappa^o (\mu / 2 + B)}{2 {\kappa^o}^2 + \mu^2 + \mu B}.
\end{align}

For a more general form of stress the strain can be written as:
\begin{align}
    & \partial_x u_x = \frac{2 {\kappa^o}^2 (p_1 + p_2) - \kappa^o (v_1 + v_2) (\mu + 2 B) + \mu (\mu p_1 + B (p_1-p_2))}{(4 {\kappa^o}^2 + \mu^2) (\mu + 2 B)}, \nonumber \\
    & \partial_y u_y = \frac{2 {\kappa^o}^2 (p_1 + p_2) + \kappa^o (v_1 + v_2) (\mu + 2 B) + \mu (\mu p_2 + B (p_2-p_1))}{(4 {\kappa^o}^2 + \mu^2) (\mu + 2 B)}, \nonumber \\
    & \partial_x u_y = \frac{\kappa^o (p_1 -p_2)}{4 {\kappa^o }^2 + \mu^2} - \frac{\theta + v_1 - v_2}{\varepsilon} + \frac{v_2 - v_1}{\kappa^c} + \frac{\mu (v_1 + v_2)}{2 (4 {\kappa^o}^2 + \mu^2)}, \nonumber \\
    & \partial_y u_x = \frac{\kappa^o (p_1 -p_2)}{4 {\kappa^o }^2 + \mu^2} + \frac{\theta + v_1 - v_2}{\varepsilon} + \frac{v_1 - v_2}{\kappa^c} + \frac{\mu (v_1 + v_2)}{2 (4 {\kappa^o}^2 + \mu^2)}, \nonumber \\
    & \phi = -\frac{\theta + v_1 - v_2}{\varepsilon}.
\end{align}

\section{Dispersion relation from cubic equation}

The dispersion relation (Note, that frequency $\omega = i \Lambda$) of an odd-Cosserat solid can be obtained as a solution to the cubic equation :
\begin{align}
    \Lambda^3 + b~\Lambda^2 + c~\Lambda + d = 0,
\end{align}
where,
\begin{align}
    & b = 4 \kappa^c + \kappa^c k^2 + 8 \mu k^2, \nonumber \\
    & c = 32 \mu \kappa^c k^2 + 4 k^4 (4 \mu^2 + \mu \kappa^c + 4 {\kappa^o}^2) \nonumber \\
    & d = 64 \kappa^c k^4 ({\kappa^o}^2 + \mu^2).
\end{align}

The general solutions of the equation is given by:
\begin{align}
    \Lambda = & -\frac{b}{3} - \frac{2^{1/3} (-b^2 + 3c)}{3 \Theta} + \frac{\Theta}{3 \times 2^{1/3}}, \nonumber \\
    & -\frac{b}{3} + \frac{(1 + i \sqrt{3})(-b^2 + 3c )}{3 \times 2^{2/3} \Theta} - \frac{(1 - i \sqrt{3}) \Theta}{6 \times 2^{1/3}}, \nonumber \\
    & -\frac{b}{3} + \frac{(1 - i \sqrt{3})(-b^2 + 3c )}{3 \times 2^{2/3} \Theta} - \frac{(1 + i \sqrt{3}) \Theta}{6 \times 2^{1/3}}.
\end{align}
where,
\begin{align}
    \Theta = \left( -2b^3 + 9 b c - 27 d +3 \sqrt{3} \sqrt{-b^2 c^2 + 4 c^3 + 4 b^2 d - 18 b c d + 27 d^2} \right)^{1/3}
\end{align}

In the case of $\kappa^o = 0$, the above relations greatly simplify and we obtain:
\begin{align}
    \Lambda = & - 4 k^2 \mu, \nonumber \\
    & -4 \kappa^c - \kappa^c k^2 - 4 \mu k^2 \pm \sqrt{-64 k^2 \kappa^c \mu + ((4 + k^2) \kappa^c + 4 k^2 \mu)^2}.
\end{align}

\section{Rayleigh Edge modes}

Similar to the bulk waves discussed above one can derive surface waves in the above system of equations. We consider the equations:
\begin{align}
    \dot{\bf u} = \mu \nabla^2 {\bf u} + \kappa^o \nabla^2 {\bf u}^* - \frac{\kappa^c}{4} \nabla^* (\nabla \times {\bf u}),
\end{align}
and an ansatz ${\bf u} = {\bf U} e^{i(kx - \omega t)} e^{a y}$ in a plane semi-infinite in the $y$--direction and consider $y<0$. This gives a dispersion relation given by:
\begin{align}
    &\omega = \frac{i}{8} (a^2 - k^2)
    \left[ 8 \mu + \kappa^c \pm \left({\kappa^c}^2 - 64 {\kappa^o}^2  \right) ^{1/2} \right]
\end{align}

Let us now consider zero normal stress boundary conditions ($\sigma_{yy} = 0$ and $\sigma_{xy} = 0$) we obtain the following as boundary conditions at $y=0$ :
\begin{align}
    & \sigma_{yy} = 0 = B \partial_x u_x + (\mu + B) \partial_y u_y - \kappa^o \partial_x u_y - \kappa^o \partial_y u_x, \nonumber \\
    & \sigma_{xy} = 0 = \kappa^o \partial_y u_y - \kappa^o \partial_x u_x + \frac{\kappa^c}{2} \phi + \left( \frac{\mu}{2} - \frac{\kappa^c}{2} \right) \partial_x u_y + \left( \frac{\mu}{2} + \frac{\kappa^c}{2} \right) \partial_y u_x.
\end{align}
For our purpose, here we will set $\phi=0$ as a boundary condition on $\phi$ at $y= 0$. Now the field ${\bf u}$ will have a transverse and a longitudinal component such that $\nabla \cdot {\bf u}^t = 0$ and $\nabla \times {\bf u}^l = 0$, the superscripts $l$ and $t$ denote longitudinal and transverse respectively. We will also use the notation $a_l$ and $a_t$ to denote the decay length of the longitudinal and transverse waves respectively and $U_l$ and $U_t$ for amplitudes of the longitudinal and transverse waves. Therefore, we obtain:
\begin{align}
    & u^t_x = a_t U_t \exp(i k x + a_t y - i \omega t), \nonumber \\
    & u^t_y = -i k U_t \exp(i k x + a_t y - i \omega t), \nonumber \\
    & u^l_x = k U_l \exp(i k x + a_l y - i \omega t), \nonumber \\
    & u^l_y = -i a_l U_l \exp(i k x + a_l y - i \omega t).
\end{align}
Now, we can write the $x$ and $y$ components in terms of the longitudinal and transverse components. Therefore, we get:
\begin{align}
    & u_x = (a_t U_t e^{a_t y} + k U_l e^{a_l y}) \exp(i(kx - \omega t)), \nonumber \\
    & u_y = -i (k U_t e^{a_t y} + a_l U_l e^{a_l y}) \exp(i(kx - \omega t)).
\end{align}
Using the above forms we can compute the spatial derivatives of ${\bf u}$ which is:
\begin{align}
    & \partial_x u_x = i k (a_t U_t e^{a_t y} + k U_l e^{a_l y}) \exp(i(kx - \omega t)), \nonumber \\
    & \partial_x u_y = k (k U_t e^{a_t y} + a_l U_l e^{a_l y}) \exp(i(kx - \omega t)), \nonumber \\
    & \partial_y u_x = (a_t^2 U_t e^{a_t y} + k a_l U_l e^{a_l y}) \exp(i(kx - \omega t)), \nonumber \\
    & \partial_x u_y = -i (k a_k U_t e^{a_t y} + a_l^2 U_l e^{a_l y}) \exp(i(kx - \omega t)).
\end{align}

At $y=0$ we get:

\begin{align}
    & \partial_x u_x = i k (a_t U_t + k U_l) \exp(i(kx - \omega t)), \nonumber \\
    & \partial_x u_y = k (k U_t + a_l U_l) \exp(i(kx - \omega t)), \nonumber \\
    & \partial_y u_x = (a_t^2 U_t + k a_l U_l) \exp(i(kx - \omega t)), \nonumber \\
    & \partial_x u_y = -i (k a_k U_t + a_l^2 U_l) \exp(i(kx - \omega t)).
\end{align}

Now, going back to the condition on normal stress we obtain:
\begin{align}
   & U_t \left[ -i 2k \kappa^o a_t + \frac{\mu}{2} (k^2 + a_t^2) -\frac{\kappa^c}{4} (k^2 - a_t^2) \right] + U_l \left[ -i \kappa^o (k^2 + a_t^2)  + \mu k a_l \right] = 0, \nonumber \\
   & U_t \left[ -i \mu k a_t - \kappa^o (k^2 + a_t^2) \right] + U_l \left[ i B (k^2 - a_l^2) -i \mu a_l^2 -2 \kappa^o k a_l\right] = 0
\end{align}

While, Cosserat elasticity has an effect of rescaling parameters in the Rayleigh waves the presence of odd elasticity introduces additional phase in the amplitudes indicating that the Rayleigh surface waves are circularly polarised and the sense of polarisation given by the sign of odd elasticity.

\end{widetext}


%

\end{document}